\documentclass[conference, 10pt]{IEEEtran}

\ifCLASSINFOpdf

\else

\fi
\usepackage{graphicx}
\usepackage{multirow}
\usepackage{amsmath}
\usepackage{url}
\usepackage{color}

\begin{document}
%
\title{SEED: Public Energy and Environment Dataset for Optimizing HVAC Operation in Subway Stations}

\author{\IEEEauthorblockN{Yongcai Wang, \emph{Member IEEE}}
\IEEEauthorblockA{Institute for Interdisciplinary \\ Information Sciences (IIIS) \\
Tsinghua University, \\ Beijing,  P. R. China, 100084\\
 wangyc@tsinghua.edu.cn}
\and
\IEEEauthorblockN{Haoran Feng}
\IEEEauthorblockA{National Engineering Research \\Center of Software Engineering, \\ Peking University, \\ Beijing, P. R. China, 100084\\
haoran@pku.edu.cn}
\and
\IEEEauthorblockN{Xiao Qi}
\IEEEauthorblockA{Institute for Interdisciplinary \\Information  Sciences (IIIS) \\
Tsinghua University, \\ Beijing,  P. R. China, 100084\\
qixiao0113@gmail.com}
}

\maketitle

\begin{abstract}
For sustainability and energy saving, the problem to optimize the control of heating, ventilating, and air-conditioning (HVAC) systems has attracted great attentions, but analyzing the signatures of thermal environments and HVAC systems and the evaluation of the optimization policies has encountered inefficiency and inconvenient problems  due to the lack of public dataset. In this paper, we present the Subway station Energy and Environment Dataset (SEED), which was collected from a line of Beijing subway stations, providing minute-resolution data regarding the environment dynamics (temperature, humidity, CO2, etc.) working states and energy consumptions of the HVAC systems (ventilators, refrigerators, pumps),  and hour-resolution data of passenger flows.  We describe the sensor deployments and the HVAC systems for data collection and for environment control, and also present initial investigation for the energy disaggregation of HVAC system, the signatures of the thermal load,  cooling supply, and the passenger flow using the dataset.

\end{abstract}

\IEEEpeerreviewmaketitle

\section{Introduction}
For low-carbon, sustainability and environment friendly living, reducing the energy consumptions of electrical appliances has attracted great attentions, among which, optimizing the operations of Heating Ventilation and Air Conditioning (HVAC) systems plays a major role, because the HVAC systems are energy consuming giants in our living environments. For example,  the HVAC systems in a commercial building may consume nearly 50\% of overall energy \cite{perez2008review} and the HVAC system in a subway station can consume more than 40\% of the total power \cite{lu_analysis_2011}.  If we can decrease the energy consumption of the HVAC system a few percents, for example 10\%, dramatical energy can be saved.  

A major way to save energy for the HVAC systems  is to design optimal control strategies to minimize the overall energy consumption while still maintaining the satisfied indoor thermal comfort and healthy environment \cite{wang2008supervisory}.  This process generally needs three procedures: 1) identifying the load signatures of the buildings and the cooling-energy patterns of the HVAC systems; 2) designing the optimal control policies; 3) evaluating the control policies. Current approaches generally used simulation, or model-based methods to tackle the diversity and complexity of thermal exchanging in different kinds of buildings. Because the buildings' surfaces and structures are diverse and the inner states of the HVAC system are complex to monitor, it is generally expensive to build reasonable models, and at least in some extend lacks fidelity in design and evaluation. 

On the other hand, although it is highly relevant to use data mining or machine learning techniques to identify the signatures of the thermal environments and the HVAC systems, which are also powerful tools for   optimizing the control policies, very few work has been seen in this area. 
It is at least partially due to lack of publicly available dataset in this domain, which is mainly because of the difficulty for monitoring the dynamics of the thermal environments, user states, and the generally non-accessing of the HVAC working states. Also in the interdisciplinary areas of power and computing, in a very closed domain, some recent published data set: REDD\cite{kolter2011redd}, BLUED\cite{anderson2012blued}, Smart*\cite{barker2012smart} have dramatically benefited the studies in energy disaggregation in smart homes. However, there are still few dataset regarding the real, long-term, fine-grained working states, thermal environment conditions and user states in HVAC systems.

In this paper, we present the Subway station Energy and Environment Dataset (SEED), which was collected in August and September in the summer of 2013, over multiple stations from a line of Beijing subway. It  provides minute-resolution, comprehensive data regarding the environment dynamics (temperature, humidity, CO2, etc.), working states and energy consumptions of the HVAC systems (ventilators, refrigerators, pumps), and hour-resolution data of passenger flows. These data was a part of the data measured and recorded during our projects for developing the autonomous HVAC control systems for Beijing metro stations. For the sake of protecting the privacy of the subway stations, the name of the stations and the lines are hidden in the public dataset, which will not affect its usage. We describe the deployment of sensors, the HVAC systems, the hardware and the software platform for environment, HVAC states, and the passenger flow monitoring. We also present initial investigation for the energy disaggregation, environment and passenger load signature analysis and HVAC cooling supply signature analysis utilizing the dataset.  The entire dataset and the notes to explain it are available online at: \url{http://iiis.tsinghua.edu.cn/~yongcai/SEED/}.

The remainder of this paper is organized as following. Background and related works are introduced in Section II.  Sensor deployments and the system architecture of the subway HVAC system are introduced in  Section III.  The overview of the dataset and some attributes are highlighted in Section IV.  We present basic investigations on the power disaggregation, signatures on the loads and cooling supply in Section V.  Conclusion and further works are presented in Section VI.




\section{Related Work and Background}
\subsection{Optimization of HVAC systems}
HVAC system optimization generally includes three steps:
\subsubsection{Signature Identification} which is to identify the signatures of thermal environments and the HVAC systems. This step generally need extensive in-field survey, measurements under controlled HVAC operations, and  post data processing.  In case the in-field measurements are infeasible because of lacking the real systems or resources to conduct measurements, theoretical models or simulation based models are used instead.  The most widely used simulators include DeST\cite{yan2008dest} and EnergyPlus\cite{crawley2000energy}, which provide detailed models to simulate the thermal exchanging patterns in different kinds of buildings. The encoded parameters of buildings include the size, surface styles, thickness, materials of walls, roofs, windows, doors, and a lot of other parameters, so that it is generally time consuming to setup an acceptable simulation model. In theoretical model aspect,  dynamic model of an HVAC system for control analysis was presented in \cite{tashtoush_dynamic_2005}. The authors proposed to use Ziegler-Nichols rule to tune the parameters to optimize PID controller.  Multi agent-based simulation models were studied in \cite{andrews_designing_2011} to investigate the performance of HVAC system when occupants are participating. More simulation models and theoretical models can be referred to survey in \cite{trvcka2010overview}. 
\subsubsection{Designing Optimal Control Policy} is to design adaptive control strategies  or the optimal setpoints based on rough theoretical or simulation-based models to minimize the overall energy consumption of the HVAC system while still maintaining the required indoor thermal comfort.  Tremendous research efforts have been devoted in this area,  especially for sustainable buildings \cite{kelman_analysis_2011}\cite{house_optimal_1995}\cite{yang_optimal_2012}. Various optimization techniques have been exploited in existing studies, including evolutionary computing\cite{fong_hvac_2006}, genetic algorithm and neutral networks etc \cite{chow2002global}. A survey of the optimization methods  was conducted by \cite{wang2008supervisory}. 
\subsubsection{Evaluate the Control Policy} Since the HVAC systems are running in practical environments, it is generally infeasible to directly test the immature control policies in the HVAC systems. Therefore, most of the control policies are evaluated via simulations in their design phase, which in some extend lacks the fidelity of system dynamics.  By providing public, fine-grained dataset regarding thermal environments and HVAC system energy and state logs, all the above three steps can be benefited.  

\subsection{Optimizing HVAC Systems in Subway Stations}
As a branch of HVAC systems for large buildings, the HVAC systems in subway stations have also attracted great attentions. One of the most closely related work is the SEAM4US (Sustainable Energy mAnageMent for Underground Stations) project established in 2011 in Europe\cite{seam4us}. It studies the metro station energy saving mainly from the modeling and controlling aspect. Multi-agent and hybrid models were proposed in\cite{serban_common_2012,roberta_ansuini_hybrid_2012}, and adaptive and predictive control schemes were proposed for controlling ventilation subsystems to save energy \cite{giretti_energy_2012}.  
Another related work reported the factors affecting the range of heat transfer in subways \cite{hu_numerical_2008}.  They showed by numerical analysis that how the heat was transferred in tunnels and stations. Reference \cite{awad_environmental_2002} studied the environmental characters in the subway metro stations in Cairo, Egypt, which showed the different environment characters in the tunnel and on the surface. 
\subsection{Related Datasets}
This paper focuses on providing public dataset for efficiency and convenience in studying the HVAC optimization problems. Although few datasets are available in HVAC studies, a series of public datasets were published recently in the area of energy disaggregation in smart homes, including REDD\cite{kolter2011redd}, BLUED\cite{anderson2012blued}, Smart*\cite{barker2012smart} etc.  The prevalence of these datasets has strongly benefited the application of machine learning methods into energy disaggregation area. For the related data analysis works in HVAC systems, the most related one is \cite{lu_analysis_2011}, which surveyed the energy consumption of Beijing subway lines in 2008, but without providing a dataset.

\section{Sensing and HVAC Control Systems}
The SEED dataset was constructed during our development of the autonomous HVAC energy conservation systems for Beijing subway stations.  We firstly report the deployment of sensors by using a subway station as an example. 

\subsection{Sensor Deployment}
\begin{figure}[h]
\begin{center}
\includegraphics[width=3in]{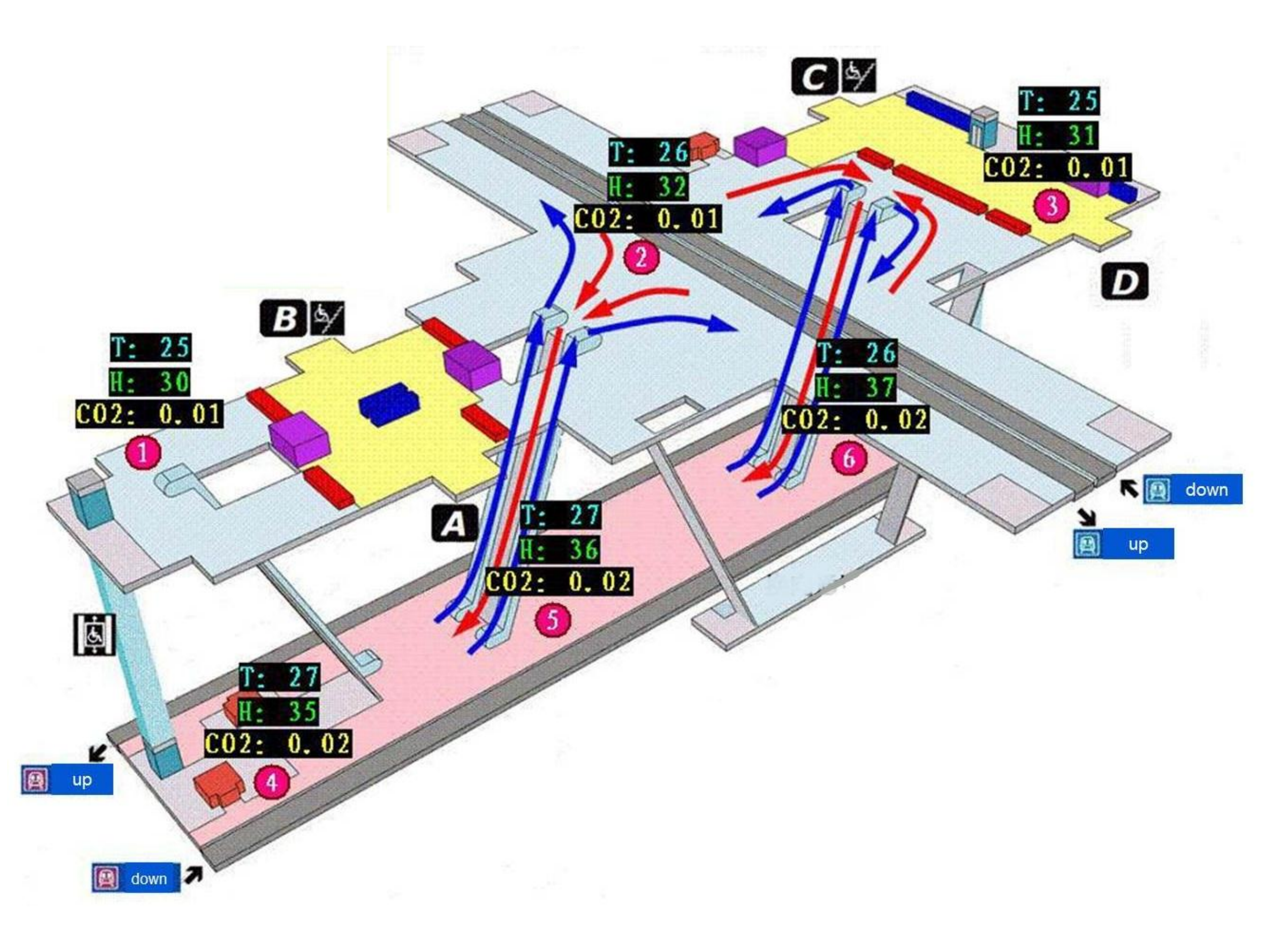}
\caption{The structure of  subway station A and the deployment of sensors for environment monitoring}
\label{subway}
\end{center}
\end{figure}

Our way to capture the thermal and the environment dynamics in the subway station is to deploy sensors to measure the indoor, outdoor temperatures, passenger flows and power consumptions of the HVAC systems in real-time. In  subway station A (we hide the name for the sake of privacy protection), which is a transferring station between two lines in Beijing subway, we deployed different kinds of sensors and smart meters to measure above information. The architecture of the  station and the deployment of sensors are shown in Fig. \ref{subway}, which is from a snapshot of our subway station environment monitoring interface. 
 
\subsubsection{Environment Sensors}
We deployed temperature, humidity and CO2 sensors at four points inside the subway station and two points outside the subway station to monitor the indoor and outdoor temperatures, humidity and CO2 density  respectively. The sensors are connected to a data collection server. Each sensor reports data once per minute, so the time resolution of the environment data is one-minute. The deployed positions of the sensors in the Station A are shown in Fig.\ref{subway}.  Similar sensor deployment and data collection strategy are also used in other stations in the same line to collect the environment data in real-time. 
\subsubsection{Passenger Flow} 
Since the thermal brought in by the passengers is also an important source of heat, we acquired the passenger flow data from the operating company of the subway. The passenger flow was recorded by the ticket checking system. In SEED data set, passenger flows over multiple days in multiple stations are provided. We will compare the different temporary patterns of the passenger flows in the working days and in the weekends in the next section.  
\subsubsection{Run-time Parameters and States of the HVAC System}
By deploying power meters, sensors, and by readings from the internal sensors of the HVAC system,  the run-time parameters and working states of the HVAC system, including the data of the refrigerators, ventilators, cooling towers, pumps and the valves of the HVAC system are monitored. The HVAC systems in different subway stations have the same architecture. Each HVAC system in a station contains 3 refrigerators, 2 supply fans, 2 return fans, 2 exhaust fans, 4 cooling pumps, 4 chilling puns, a set of valves. The sensor readings of these devices are listed in Table \ref{table3}. These data is reported to the central data collection server in one-minute time resolution.

\begin{table}[htdp]
\caption{List of data types provided in SEED dataset}
\begin{center}
\begin{tabular}{|p{0.9in}|p{1.35in}|p{0.6in}|} \hline 
\textbf{Environment Info} & \textbf{Type of Sensors} & \textbf{Type of Values}\\ \hline
6 Temperature sensors & Temperature at $i$th outdoor sensor & ${}^{o}$C  \\\cline{2-3}
 & Temperature at $i$th indoor sensor & ${}^{o}$C  \\\hline 
6 Humidity sensors & Humidity at $i$th indoor sensor& \%  \\\cline{2-3} 
& Humidity at $i$th outdoor sensor& \%  \\\hline
6 CO2 sensors & CO2 at $i$th outdoor sensor & mg/kg  \\\cline{2-3} 
& CO2 at $i$th indoor sensor&  mg/kg  \\\hline
\textbf{Devices of HVAC} & \textbf{Parameters or States} & \textbf{Types of Value} \\ \hline 
3 Refrigerators & Power of $i$th Refrigerator & Watt \\ \cline{2-3}
 & Current of $i$th Refrigerator & Ampere \\ \cline{2-3}
 & State of $i$th Refrigerator & 0/1 \\ \cline{2-3}
 & Cool Water Temperature  & ${}^{o}$C \\ \cline{2-3}
 & Return Water Temperature  & ${}^{o}$C \\ \hline
2 Supply fans  & Power of $i$th fan & Watt \\ \cline{2-3}
 2 Return fan & Current of $i$th fan & Ampere \\ \cline{2-3}
 2 Exhaust fan & Working State of of $i$th fan & 0/1 \\ \cline{2-3}
 & Supply air temperature & ${}^{o}$C \\ \cline{2-3}
 & Return air temperature & ${}^{o}$C \\ \cline{2-3}
 & Exaust air temperature & ${}^{o}$C \\ \hline 
4 Cooling pumps & Power of $i$th pump & Watt \\ \cline{2-3} 
 4 Chilling pumps & Current of $i$th pump & Ampere \\ \cline{2-3} 
 & Working State of $i$th pump  & 0/1 \\ \hline 
Valves & States of $i$th valve & \% \\ \hline 
Events & Operating logs & time + event \\\hline
Frequency changers & Logs of frequency changers & time+ event\\\hline  
\textbf{Passenger Flow} & \textbf{Data type} & \textbf{Types of Value} \\ \hline 
& number of checked in passengers & $n$/hour \\ \cline{2-3}
& number of checked out passengers & $n$/hour \\ \hline
\end{tabular}\end{center}
\label{table3}
\end{table}%
\subsection{Hardware and Software of HVAC Monitoring and Control}
\subsubsection{Hardware}
\begin{figure}[h]
\begin{center}
\includegraphics[width=2.5in]{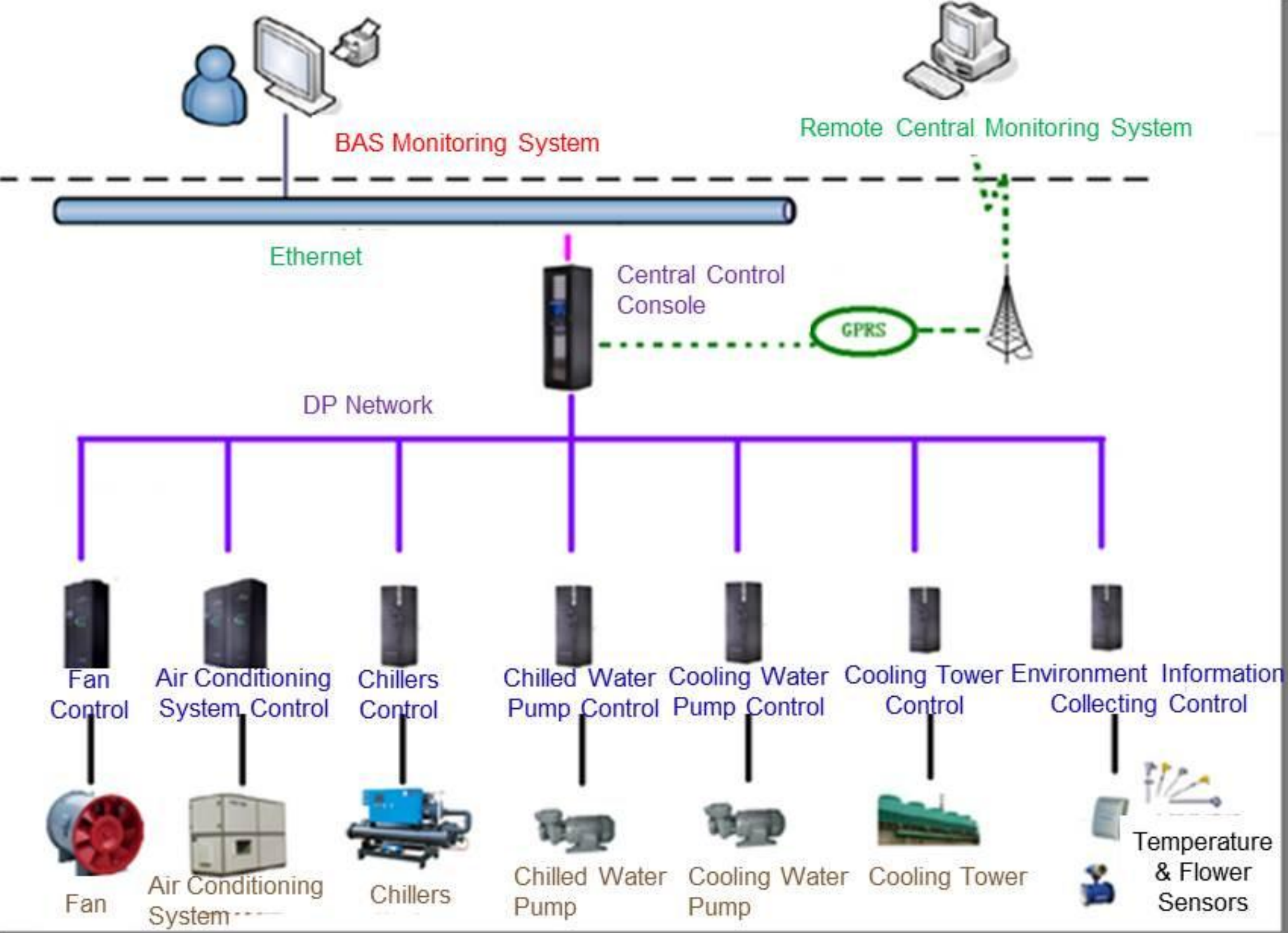}
\caption{Architecture of HVAC system in a subway station}
\label{system}
\end{center}
\end{figure}

The SEED dataset contains data types in above list collected from three stations over multiple days.  The HVAC systems in different stations shares the same structure, which is illustrated in Fig.\ref{system}. The devices in the HVAC systems are all from Carrier \url{http://www.carrier.com.cn}.  We added  the new air temperature sensors, return air temperature sensors and deployed Profibus DP network to connected the sensors into the information collection server. We have also developed the control cabinet and autonomous control logics for the HVAC system. 
All data are collected by the information collection server to be reported to the central control console in real-time. 
 \subsubsection{Software} Based on the data collected in real-time by the deployed sensors, both the environment monitoring system and the HVAC working state monitoring systems were developed.  Fig.\ref{subway} shows the snapshot of environment monitoring interface.  

The overall sensing and control systems were established in the spring of 2013 and they have run during the whole summer of 2013. In the SEED data set, we chose data from August and September, including data from both the very hot days and data for the days when outdoor temperatures are lower in indoor temperatures. 

\begin{figure}[h]
\begin{minipage}{0.48\linewidth}
\centering
\includegraphics[width=1.65in]{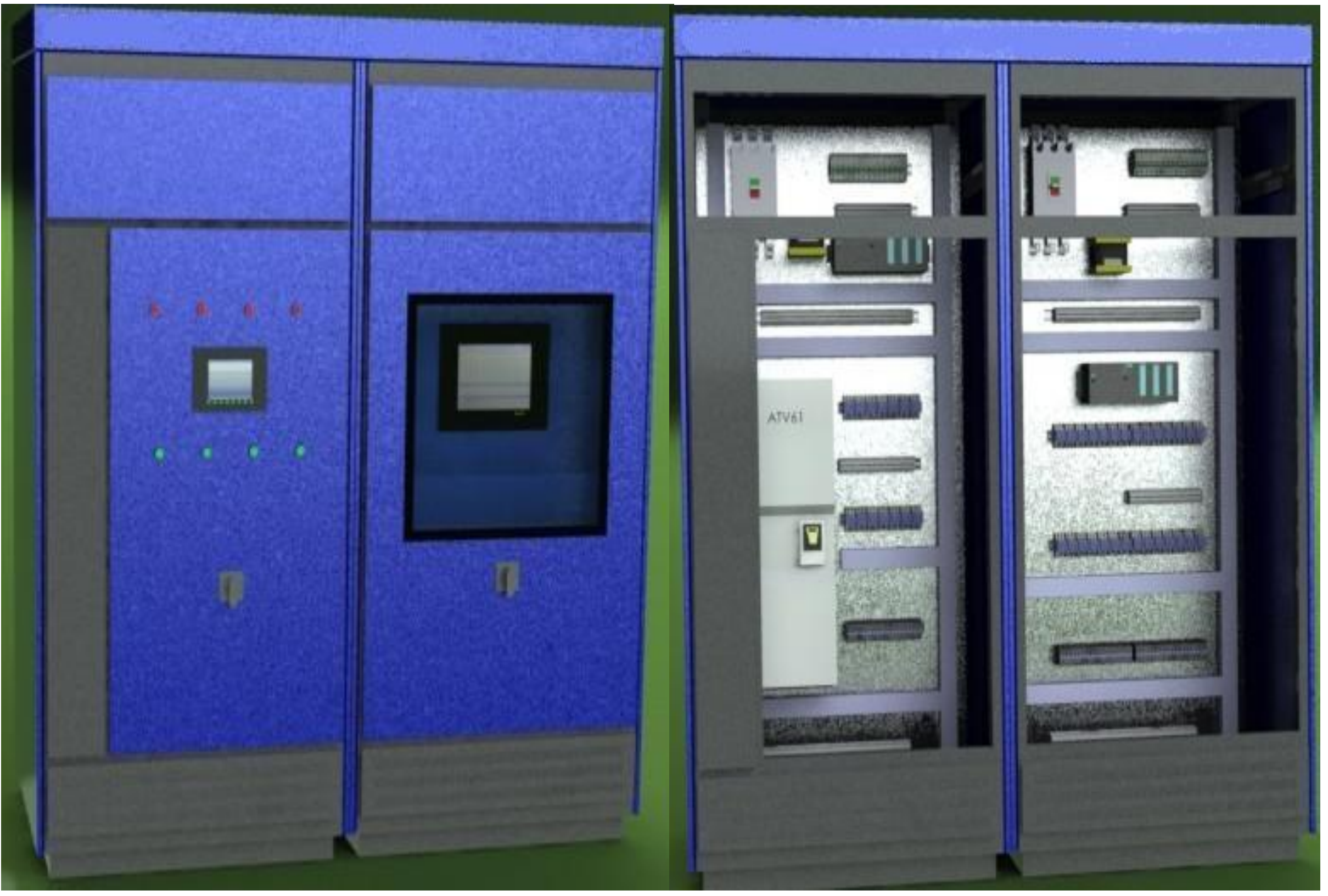}
\caption{Hardware of data collection server and control cabinet}
\label{hardware}
\end{minipage}%
\begin{minipage}{0.48\linewidth}
\centering
\includegraphics[width=1.5in]{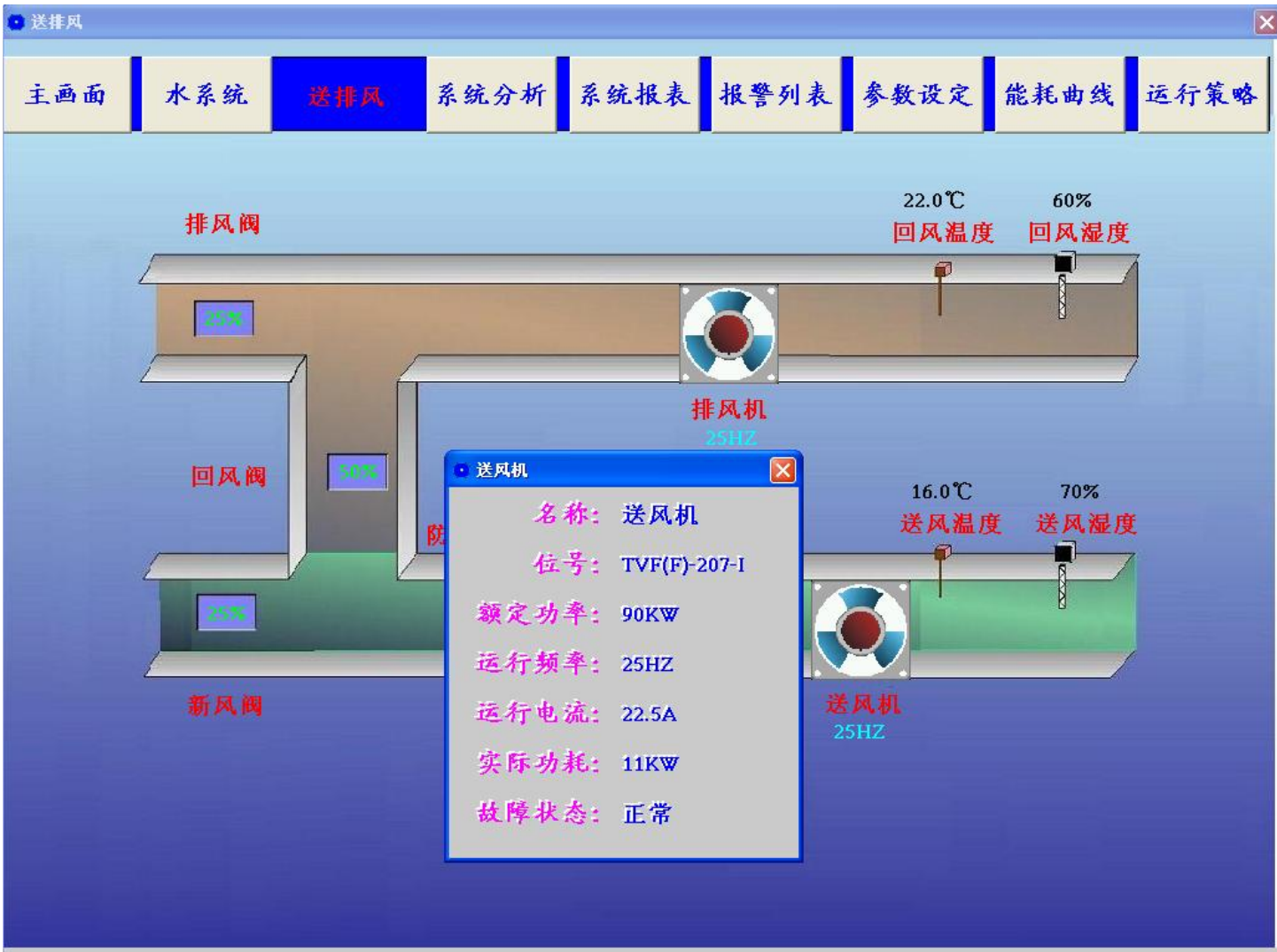}
\caption{Interface of HVAC working state monitoring software}
\label{software}
\end{minipage}%
\end{figure}
\section{Basic Investigation to SEED Dataset}
We conducted basic research on the SEED dataset to investigate basic features of the thermal environments in the subway stations and the features of the HVAC system shown by the data. 
\begin{enumerate}
\item Energy disaggregation in the HVAC system.
\item Temperature difference Vs. States of Refrigerator. 
\item Signatures of the passenger flow.
\item Correlation features of CO2 and passenger flow.
\item Responding speed to cooling supply;   
\end{enumerate}
\subsection{Energy Disaggregation in HVAC}
Because the HVAC system is an energy consuming giant, to understand how the energy was consumed in the HVAC system is of the primary interests to many researchers. We select data from 8.21 - 8.23, 8.29 - 8.31, and 9.1 -9.30 three periods from a subway station to investigate the disaggregated energy consumption in the HVAC systems (other stations have similar features), when the outdoor temperatures are different.  The average peak temperature of these three periods are $35^{o}C$, $31^{o}$C and $27^{o}$C respectively. 

Fig. 5 shows the daily average energy consumptions of the refrigerator, chilled pump, cooling pump and fans in three periods. Some interesting phenomena can be seen: 
1)  The energy consumptions of the HVAC are highly relevant to the outdoor temperatures. \emph{The higher is the daily average outdoor temperature, the higher is the daily energy consumption.} 
2)  The fans consume similar amount of energy in all three periods, so \emph{the energy differences over different periods are mainly dominated by the energy consumptions differences of the refrigerator, chilled pump and the cooling pump.}
3)  Since the pumps work only if the refrigerator is working, so their energy consumptions are strongly correlated. \emph{We can basically disaggregate the consumptions of HVAC into the consumption of ventilating (fans), which is rather stable and the consumptions of the cooling utilities (refrigerators and pumps), which are dynamic according to the outdoor weathers}.  Reducing the consumptions of the cooling utilities should be the major way for reducing consumptions of HVAC. 
\begin{figure}[h]
\begin{minipage}{0.33\linewidth}
\centering
\includegraphics[width=1.1in]{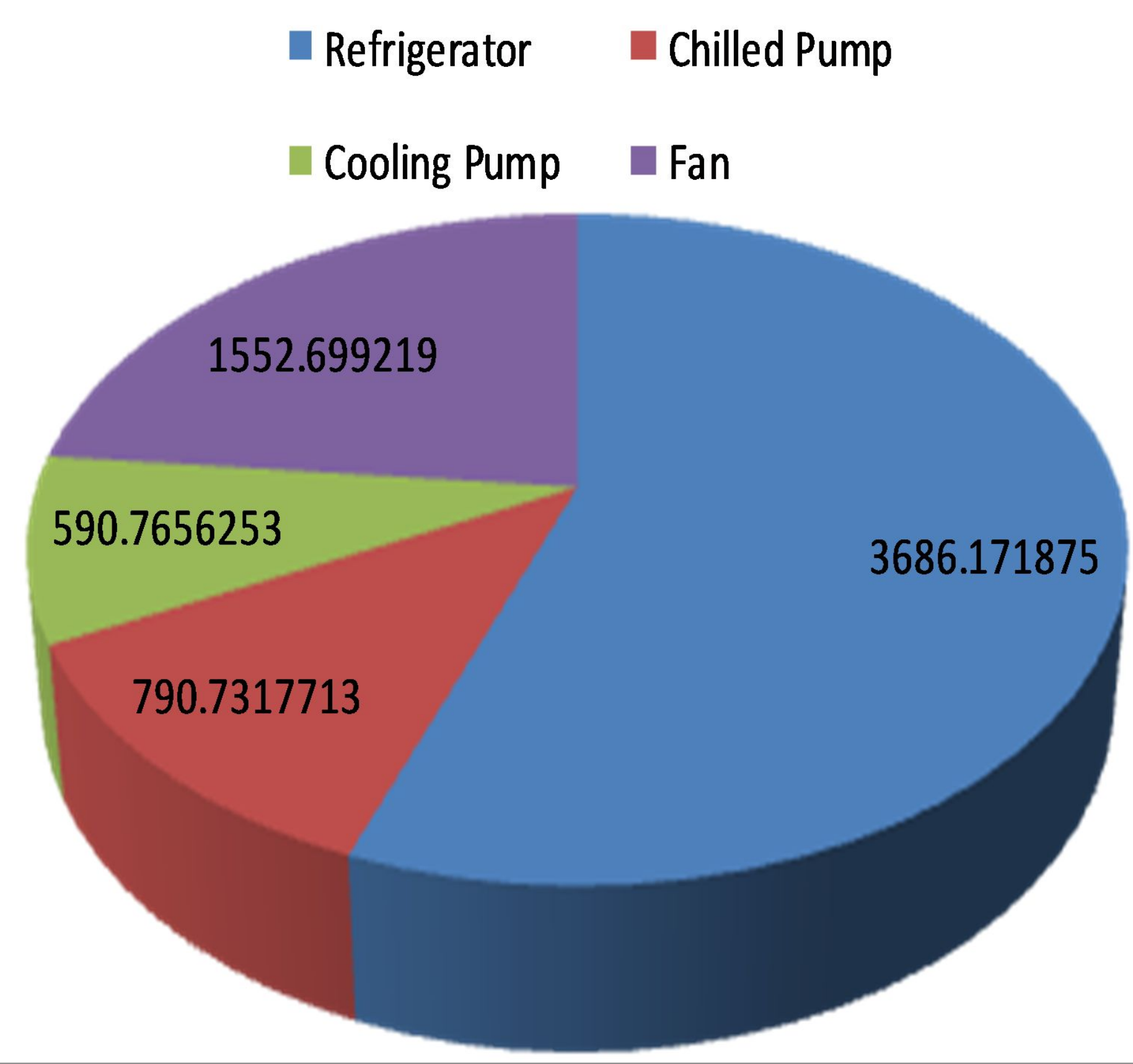}
\tiny{8.21.2013-8.23.2013}
\end{minipage}%
\begin{minipage}{0.33\linewidth}
\centering
\includegraphics[width=1.13in]{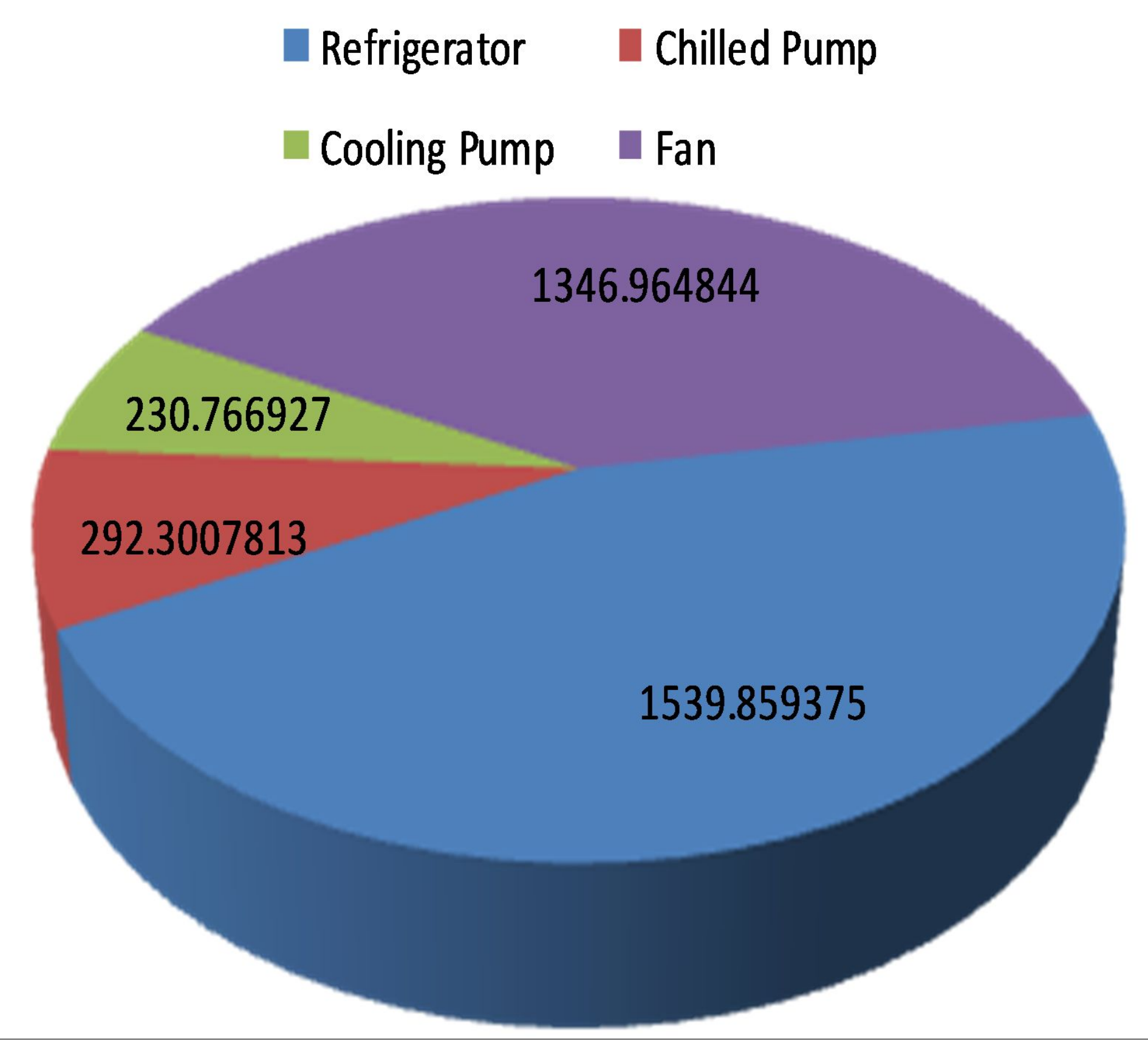}
\tiny{8.29.2013-8.31.2013}
\end{minipage}%
\begin{minipage}{0.33\linewidth}
\centering
\includegraphics[width=1.15in]{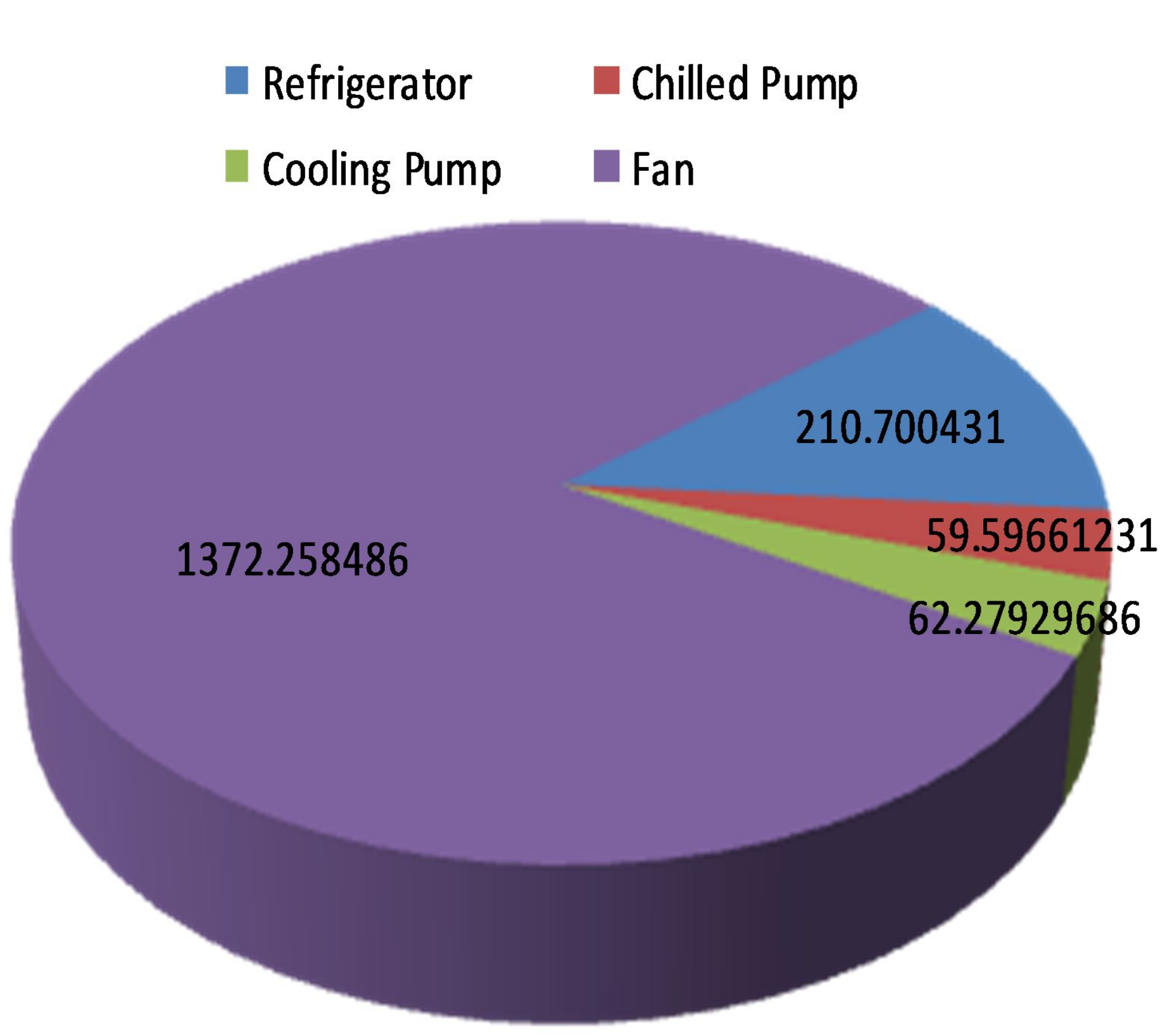}
\tiny{ 9.1.2013-9.30.2013}
\end{minipage}%
\label{disaggregation}
\caption{Comparing of average daily disaggregated energy consumptions over different time periods}
\end{figure}
 \begin{figure}[b]
\begin{minipage}{0.48\linewidth}
\centering
\includegraphics[width=1.65in]{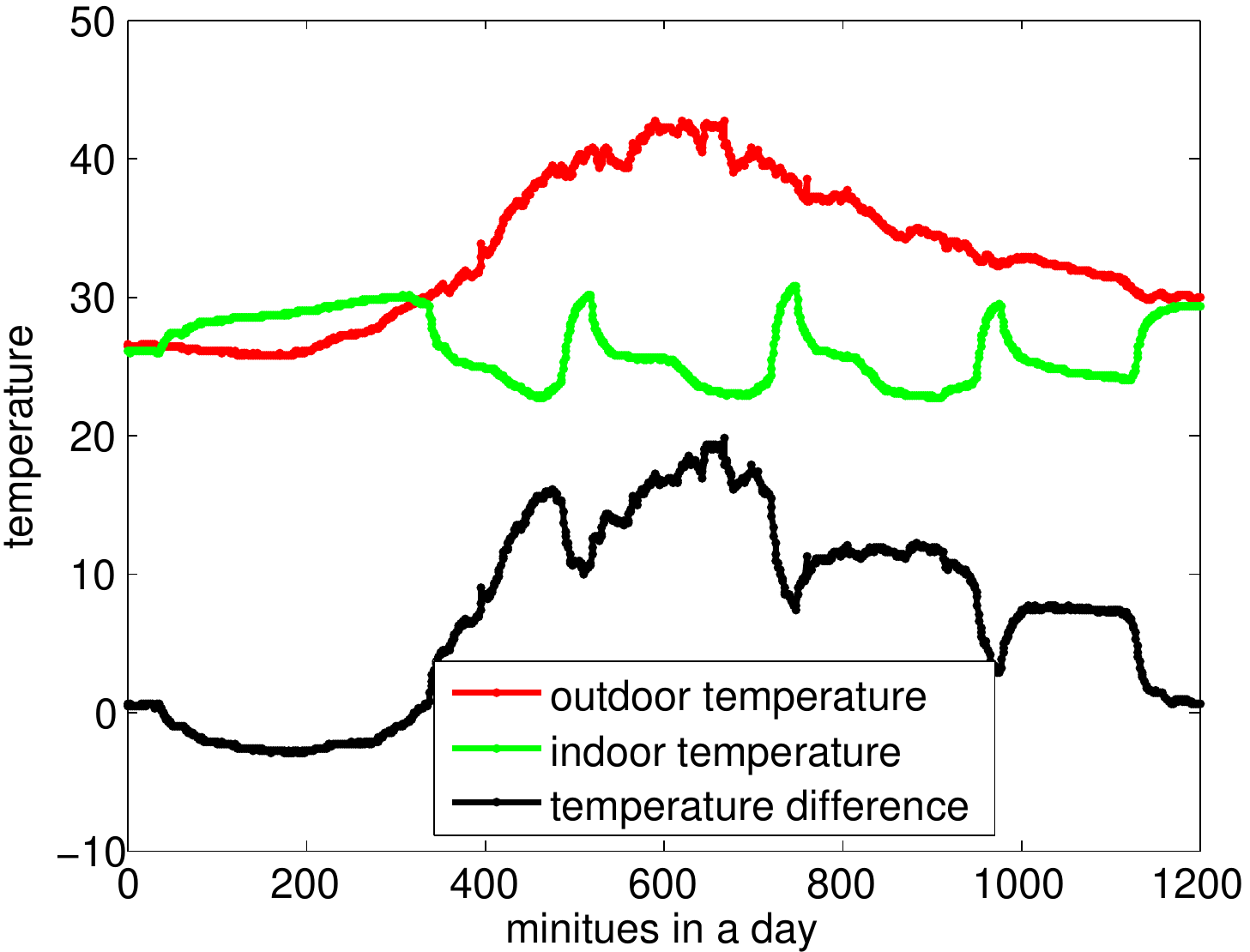}
\tiny{a) Indoor, outdoor temperature variations}
\end{minipage}%
\begin{minipage}{0.48\linewidth}
\centering
\includegraphics[width=1.65in]{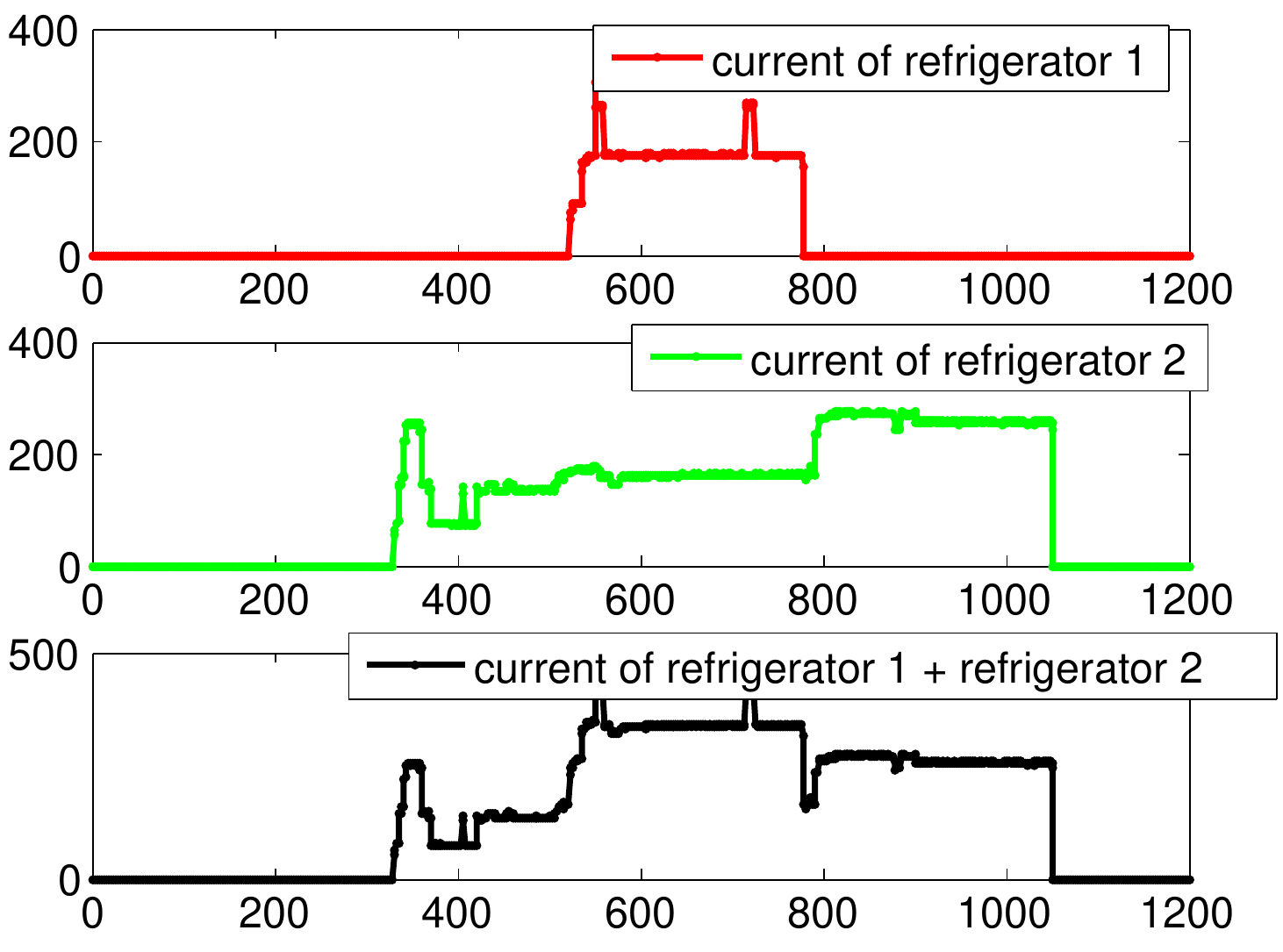}
\tiny{b) States and currents of the refrigerators}
\end{minipage}%
\caption{Indoor outdoor temperature differences VS. the states and the consumptions of the refrigerators. }
\label{response}
\end{figure}

 \subsection{Temperature Differences VS. States of Refrigerators}
To control the indoor temperature at the desired temperature point, the  refrigerators and the pumps work adaptively to response to the temperature variations. We investigated via SEED dataset how the working states and energy consumptions of the refrigerators change over a day with the variations of the outdoor temperatures. Fig.\ref{response}a) shows the temperature variations and indoor-outdoor temperature differences over a day. Fig.\ref{response}b) shows the concurrent working states and energy consumptions of the two refrigerators in that day. We can basically see the working loads of the refrigerators are closely responding to the indoor-outdoor temperature differences.  
\subsection{Signature of the Passenger Flow} 
\begin{figure}[h]
\begin{center}
\includegraphics[width=3.5in]{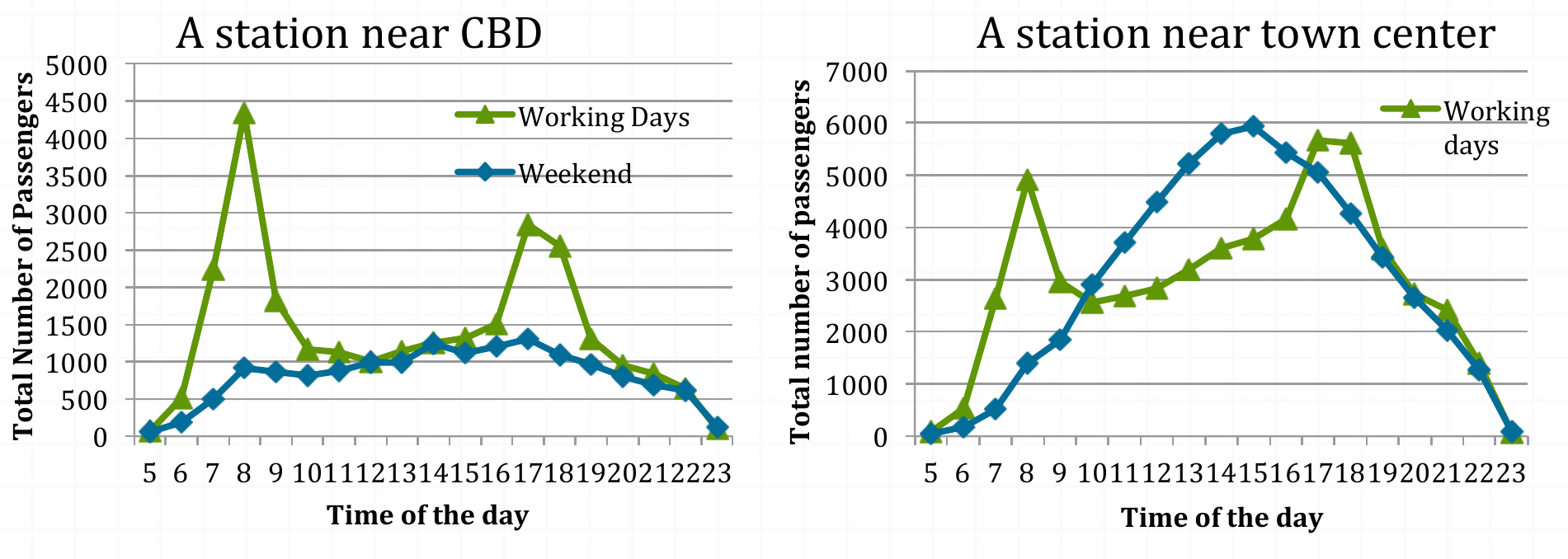}
\caption{Patterns of traffic flow are related not only to time but also to the locations of the subway stations.}
\label{flow}
\end{center}
\end{figure}

Another observation is on the passenger flow signatures.  Fig.\ref{flow} shows the patterns of passenger flows in working days and weekends of two subway stations. One station is close to CBD and the other is close to the town center. The two figures show that the signatures of the passenger flow are related not only to time but also to the locations of the stations. 

In time dimension, they show different patterns of passenger flow between the working days and the weekends. In the working days,  sharp peaks of passenger flow appear at the rush hours, while in the weekends, the passenger flow curves are different. The flow increases and decreases smoothly with peaks generally appearing at 15:00 to 16:00 pm. 

From the location dimension, for the stations close to the working places (e.g. CBD has many office towers), the peaks in rush hours in the working days are very sharp, while for locations close to leisure places (e.g. town center), the peaks in the rush hours are not very sharp. In weekends, the much less passengers go to the working places but more passengers go to the leisure places. These signatures provides hints for the smart control of HVAC system with consideration of time and location differences. 
\subsection{Correlated feature of CO2 density and the passenger flow}
\begin{figure}[h]
\begin{center}
\includegraphics[width=3.0in]{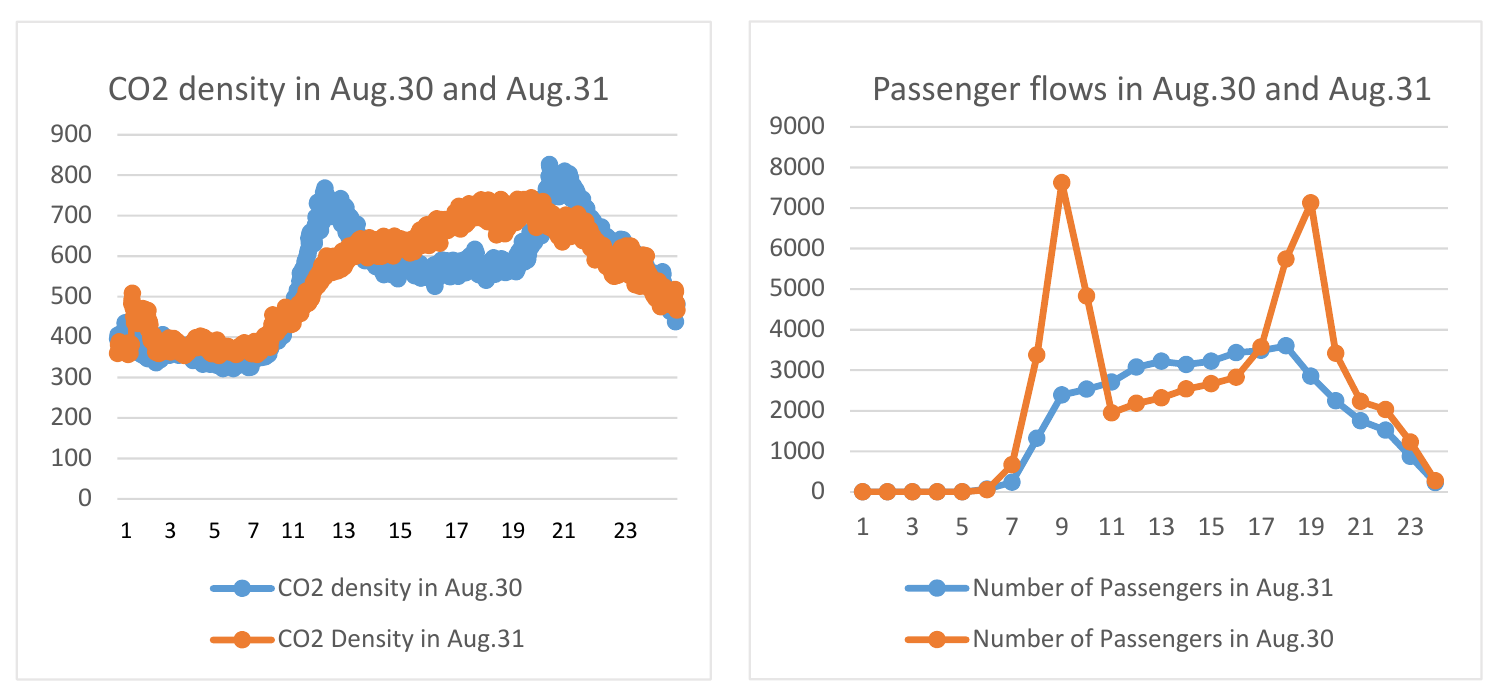}
\caption{Comparison of CO2 density and traffic flow in working days and weekends.}
\label{coflow}
\end{center}
\end{figure}

We also observed the correlated feature of CO2 density and the passenger flow over working days and weekends. It is interesting to see that the variations of CO2 density are highly relevant to the variations of passenger flows. The curves of CO2 density and passenger flow for a station in Aug. 30 (a working day) and Aug.31 (a weekend) are plotted in Fig.\ref{coflow}. The curves of the CO2 variations and the passenger flows show similar trends at corresponding time.  This result indicates that we may infer the number of passengers by the CO2 density data in case the passenger flow is not available. 
\subsection{Responding Speed of Indoor Temperature to the Cooling Supply of HVAC}
 \begin{figure}[h]
\begin{center}
\includegraphics[width=2.8in]{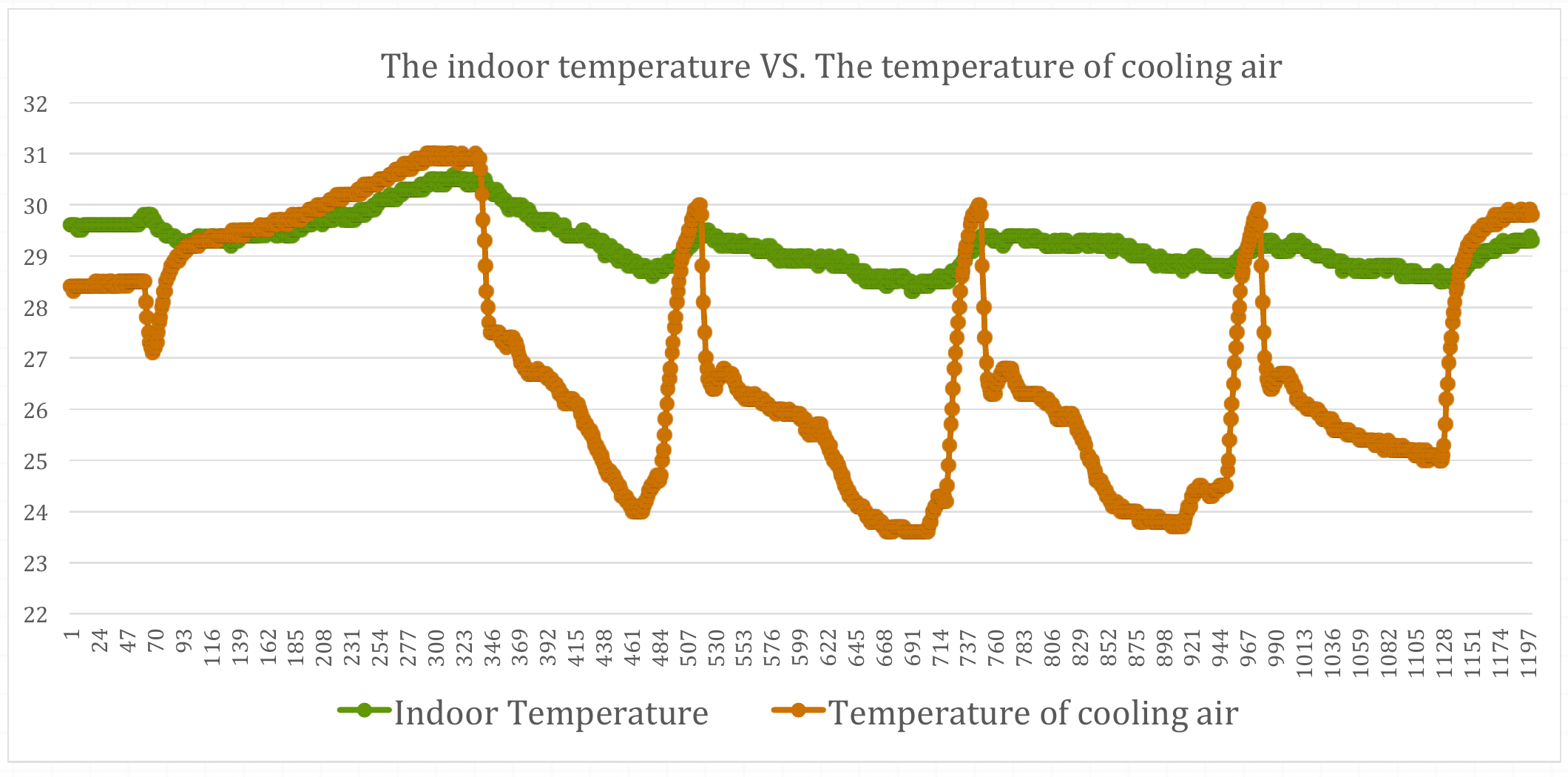}
\caption{How the indoor temperature response to the cooling supply of the HVAC system}
\label{tempfollow}
\end{center}
\end{figure}
In the last aspect, we evaluated how does the indoor temperature respond to the cooling supply of the HVAC system. We measured the cooling supply of the HVAC system by the temperature of the cooling air blowed by the cooling fans. Fig.\ref{tempfollow} shows the variation of the indoor temperature in a subway station over a day following the temperature variations of the cooling air.  We can see when the temperature of the cooling air changes, the indoor temperature changes quickly, which shows that the indoor temperature has short responding time to the cooling supply from the HVAC. It indicates that in the subway stations, the control latency is slow in the particular settings of the HVAC systems.  
\section{Conclusion and Discussion}
The paper has introduced SEED, a publicly available dataset regarding the environment, energy and working states of HVAC systems collected from multiple stations of Beijing subway over multiple days from August to September 2013. We make it publicly available for the convenience and efficiency for design and evaluation of the optimal control policies for the HVAC systems. We described the sensing and HVAC systems for data collection and environment control, and also presented our basic investigation to the energy disaggregation of HVAC, working features of the refrigerator, signatures of the passenger flow, correlation features of CO2 and the passenger flow,  and the responding speed of indoor temperature to the cooling supplies of HVAC.  
In future work, the dataset can be further investigated from different ways, such as identifying the load signatures of the subway stations, designing and evaluating the optimized control policies. 
\section*{Acknowledgment}
This work was supported by the National Basic Research Program of China Grant 2011CBA00300, 2011CBA00301, the National Natural Science Foundation of China Grant 61202360, 61033001, 61061130540, 61073174.



\bibliographystyle{abbrv}
\bibliography{seedref}
%

\end{document}